\documentclass[conference,a4paper]{IEEEtran}
\usepackage{amsmath,amssymb,amsfonts}

\usepackage{cite}
\usepackage{algorithmic}
\usepackage{graphicx}
\usepackage{textcomp}

\begin{document}

\title{ChatGPT Solves All Tested Qiskit Homework Assignments}


\author{
\IEEEauthorblockN{Alexei Kaltchenko and Gurnivaj Tiwana}
\IEEEauthorblockA{
Department of Computer Science and Physics\\
Wilfrid Laurier University\\
Waterloo, Ontario N2L~3C5, Canada\\
Email: \texttt{akaltchenko@wlu.ca}
}
}

\maketitle

\begin{abstract}
Generative artificial intelligence creates an assessment challenge in quantum software education: a student can provide a homework notebook to ChatGPT and request a completed submission. This study examined whether introductory Qiskit homework could remain autogradable while requiring students to run, review, and discuss results rather than banning AI. Three assignment packages were developed: seeded basis-state circuits with bit flips and customized measurement mappings; Quantum Fourier Transform followed by inverse-transform recovery; and seeded Deutsch-Jozsa with customized oracle masks. Their deterrence layers included deterministic personalization, non-palindromic bitstrings, varied measurement maps, simulator execution, machine-readable JSON submissions, hidden references, circuit and transpiler metrics, reflections, and optional IBM Quantum execution. For each package, one fixed student-visible assignment instance was tested in 50 separate ChatGPT sessions, producing 150 sessions overall. In every session, the final artifact was executed and passed the corresponding grader. Nine sessions, three per package, were archived with transcripts, generated files, logs, and grader outputs. No archived session required operator code changes or correction of quantum logic; eight notebook executions initially stopped at a missing optional visualization dependency and proceeded after one permitted follow-up. Under the study's operational definition, each tested instance had zero observed ChatGPT-resiliency. Analysis showed why: seeds altered parameters rather than task structure, expected results remained derivable from visible assignment logic, extensive scaffolding exposed key solution steps, and hidden grading verified output consistency without establishing independent authorship or understanding. Because the same instance was repeated within each assignment family, the results demonstrate consistent completion across sessions but do not establish solvability for every seed or possible Qiskit assessment. The findings show that the tested personalized, execution-oriented take-home designs did not prevent successful completion under the defined minimally engaged-student workflow. Correct artifacts should therefore be complemented by direct assessment of conceptual understanding through supervised modifications, oral defense, prediction, and transfer tasks.
\end{abstract}


\section{Introduction}

Hands-on programming is important in undergraduate quantum-computing education because it connects abstract state transformations and algorithms to executable circuits. Practical, software-driven courses have therefore emphasized programming exercises and projects as central learning activities \cite{mykhailova2020teaching}. The same feature now creates an assessment vulnerability. Modern large language models (LLMs) can transform natural-language specifications into source code, explanations, tests, and debugging advice. Early studies found that code-generating systems could outperform many students on introductory programming questions \cite{finnieansley2022robots}, and computing-education researchers have since called for deliberate reassessment of learning goals and assessment practices \cite{denny2024computing,kasneci2023chatgpt}.

This project began with a constructive goal: develop AI-resilient and autogradable Qiskit homework for an undergraduate quantum software development course. ``AI-resilient'' did not mean banning AI. The intended design would allow legitimate AI assistance while making a correct submission depend on a student's personalized circuit execution, verification, and interpretation. Proposed layers included deterministic seeds, simulator and hardware outputs, nontrivial measurement mappings, machine-readable answer files, hidden instructor references, circuit artifacts, and reflections tied to individual results.

Exploratory trials first showed that ordinary Qiskit exercises were highly vulnerable to ChatGPT assistance. The project then produced three AI-deterrence-modified assignment packages and associated grading artifacts. Repeated testing changed the research emphasis: all three fixed student-visible instances remained fully solvable when their text or files were provided directly to ChatGPT. The relevant adversary was not a sophisticated attacker. It was a minimally engaged student who requests a complete solution, runs the generated notebook, returns an exact error when necessary, and submits the resulting files.

The study addresses the following research question:

\emph{When ChatGPT is given the complete student-visible materials for each of three AI-deterrence-modified introductory Qiskit assignments, can it produce an executed, grader-accepted submission under a minimally engaged student workflow?}

The paper makes four contributions. First, it reports a Qiskit-specific negative result from 150 separate sessions: 50 sessions for each of three fixed assignment instances, with 150 of 150 final artifacts executing and passing the corresponding grader. Second, it evaluates complete assignment packages rather than isolated code prompts, including personalization, output formats, reflections, execution, and grading. Third, it retains an auditable subset containing frozen assignments, transcripts, generated files, logs, and grader outputs for nine sessions. Fourth, it identifies why plausible deterrence layers did not prevent completion and derives implications for human-centric engineering assessment. The title deliberately says ``all tested'' rather than all possible Qiskit homework.

\section{Related Work}

\subsection{Generative AI and Programming Assessment}

Research predating ChatGPT already showed that code-generating models could solve many introductory programming problems \cite{finnieansley2022robots}. Subsequent work has examined both productive uses of LLMs and risks to assessment validity. Denny \emph{et al.} summarize the resulting shift in computing education, where educators must reconsider what students should learn and how achievement can be evidenced when code can be generated from natural-language descriptions \cite{denny2024computing}. Kasneci \emph{et al.} similarly describe opportunities for personalized support together with risks involving accuracy, over-reliance, and assessment \cite{kasneci2023chatgpt}.

McDanel and Novak directly evaluated complete multi-part programming assignments with several LLMs and proposed strategies for making assignments more resistant \cite{mcdanel2025resistant}. Their results show that model performance depends on assignment context, starter code, tests, and task characteristics. Our work follows the same assignment-level concern but studies a specialized engineering domain. Quantum programs add circuit semantics, probabilistic measurement, qubit/classical-bit ordering, rapidly changing software interfaces, transpilation, and optional access to noisy hardware. These properties motivate deterrence mechanisms that do not arise in ordinary introductory programming, but they may also create only superficial barriers to an LLM-assisted student.

\subsection{LLMs for Quantum Code Generation}

Several recent benchmarks establish that LLMs can generate substantial amounts of quantum code. Qiskit HumanEval contains more than 100 Qiskit tasks with canonical solutions and executable tests \cite{vishwakarma2024qiskit}. QuanBench evaluates functional correctness and quantum semantic equivalence on 44 tasks and documents frequent API and circuit-logic errors \cite{guo2025quanbench}. Qiskit QuantumKatas expands the scope to 350 pedagogically structured tasks across 26 categories and 39,200 model runs \cite{cruzbenito2026quantumkatas}. QuanBench+ aligns tasks across Qiskit, Cirq, and PennyLane and shows that execution feedback can substantially improve quantum-code generation \cite{slim2026quanbenchplus}. Recent work also reports strong gains when general-purpose models are combined with iterative execution feedback rather than domain-specific fine-tuning \cite{novo2026domain}.

These benchmarks ask whether a model can generate functionally correct code for many individual tasks. Our unit of analysis is different: a complete take-home homework package as experienced by a student. A submission may include a seeded notebook, exact JSON schema, expected output, reflections, circuit statistics, and optional hardware evidence. The question is not merely whether ChatGPT implements QFT or Deutsch--Jozsa. It is whether assignment-level defenses force enough independent execution and reasoning that direct provision of the complete assignment no longer yields an executed, grader-accepted submission. This distinction positions the present work between quantum-code benchmarking and engineering-education assessment research.

\section{Study Design}

\subsection{Threat Model and Operational Measure}

The threat model is a minimally engaged student with ordinary access to ChatGPT. The student can:
\begin{enumerate}
    \item provide the assignment text, notebook, or student-visible files;
    \item request a complete solution and required submission artifacts;
    \item run the generated code in Jupyter or Colab;
    \item return an exact installation or runtime error;
    \item copy generated explanations and submit the completed work.
\end{enumerate}
The student does not receive the hidden instructor solution, hidden reference functions, expected answers, or grader feedback. This model reflects direct outsourcing rather than sophisticated prompt engineering or autonomous agents.

A session was counted as a correct completion only when the final artifact executed and passed the corresponding assignment grader after ordinary execution and permitted error feedback. For assignment package \(a\), define observed ChatGPT-resiliency as
\begin{equation}
R_a = 1-\frac{C_a}{N_a},
\end{equation}
where \(N_a\) is the number of separate ChatGPT sessions and \(C_a\) is the number yielding an executed, grader-accepted completion. This operational measure concerns completion resistance, not plagiarism detection, authorship, or learning.

\subsection{Assignment Packages}

Table~\ref{tab:assignments} summarizes the three implemented packages. Each used a deterministic synthetic or student identifier to generate a reproducible configuration. The instructor-side reference regenerated the configuration and expected output, allowing personalization without maintaining a separate answer key for every student.

\begin{table*}[t]
\caption{Implemented Qiskit assignment packages and tested deterrence layers}
\label{tab:assignments}
\centering
\footnotesize
\begin{tabular}{p{0.07\textwidth}p{0.20\textwidth}p{0.27\textwidth}p{0.36\textwidth}}
\hline
\textbf{ID} & \textbf{Quantum-programming task} & \textbf{Student work} & \textbf{AI-deterrence and grading layers} \\
\hline
HW1 & Seeded circuits, bit flips, and measurement mapping &
Prepare a personalized basis state, apply a seeded bit-flip mask, measure through a direct or reversed qubit-to-classical-bit map, run Aer, explain count-string ordering, and export \texttt{answers.json}. &
Deterministic seed; non-palindromic bit patterns; nonzero flip mask; varied measurement map; Qiskit count-string interpretation; exact JSON fields and counts; hidden deterministic reference; reflections; optional QPY export. \\
\hline
HW2 & QFT followed by inverse-QFT recovery &
Prepare a seeded input, use and inspect the supplied QFT and inverse-QFT routines, measure through a personalized map, run Aer, report circuit/transpiler metrics, explain recovery and noise, and optionally run a smaller hardware circuit. &
Seeded non-palindromic input; mapping variation; algorithm-specific circuit workflow; circuit depth and operation counts; exact JSON; hidden reference; reflections; optional QPY, backend, job, raw-count, and hardware-noise evidence. \\
\hline
HW3 & Seeded Deutsch--Jozsa &
Classify a seeded linear oracle, build the custom oracle and complete circuit, use the personalized measurement map, run Aer, report metrics and reflections, and optionally run hardware. &
Seeded constant/balanced cases; custom non-palindromic masks; classify-before-building requirement; measurement variation; exact JSON; hidden reference; circuit metrics; reflections; optional hardware evidence. \\
\hline
\end{tabular}
\end{table*}

HW1 targeted a frequent Qiskit difficulty: count keys are displayed in classical-register order, which can differ from the order in which students list qubits. HW2 moved from elementary gates to a recognizable algorithmic workflow and added circuit metrics and a hardware extension, but the student-facing notebook was heavily scaffolded. HW3 required a seed-specific oracle rather than one textbook circuit. The broader project also investigated hidden instructions and canary markers. QPY semantic validation and noisy-simulation grading were proposed, but they were not mandatory in the 150-session fixed-instance experiment and are not claimed as defeated controls.

\subsection{Repeated Fixed-Instance Trial Series}

For each assignment package, the same fixed student-visible instance was used in 50 separate ChatGPT sessions. The design therefore held the assignment content constant while resampling the model interaction across separate sessions. This tests whether successful completion was stable rather than a one-off response; it does not test generalization across many seeds or assignment variants.

All 150 final artifacts were executed and evaluated with the corresponding grader, and all 150 passed. Complete conversational and file-level provenance was retained for nine sessions, three per package. These archived sessions followed a controlled protocol: each began in a new conversation; only student-visible materials were supplied; up to two follow-ups were permitted, limited to an exact runtime error or a request for an omitted required file; and the operator supplied no conceptual hint, independent code fix, expected answer, hidden reference, or grader feedback. Transcripts, generated versions, runtime logs, final artifacts, and grader output were preserved. The nine archived sessions used the displayed ChatGPT configuration ``5.6 sol medium'' on August 16, 2026; model metadata were not retained uniformly for the other sessions.

\subsection{Historical Hardware Evidence}

Before the fixed-instance trial series, exploratory QPE, QFT, and Deutsch--Jozsa notebooks tested whether real-device execution would create a meaningful barrier. These runs are secondary evidence because they used a separate exploratory workflow. They are included to assess the project's hardware hypothesis, not in the 150-session completion count.

\section{Results}

\subsection{All 150 Final Artifacts Executed and Passed}

For each assignment package,
\begin{equation}
C_a=N_a=50,\qquad R_a=0.
\end{equation}
Thus, 150 of 150 final artifacts executed and passed the corresponding grader, and each of the three fixed instances had zero observed ChatGPT-resiliency under the operational definition. Table~\ref{tab:results} separates the full outcome count from the fully archived sessions.

\begin{table}[t]
\caption{Fixed-instance trial series and fully archived sessions}
\label{tab:results}
\centering
\footnotesize
\begin{tabular}{lccc}
\hline
 & \textbf{HW1} & \textbf{HW2} & \textbf{HW3} \\
\hline
Separate ChatGPT sessions & 50 & 50 & 50 \\
Executed final artifacts & 50/50 & 50/50 & 50/50 \\
Grader-accepted final artifacts & 50/50 & 50/50 & 50/50 \\
Observed resiliency \(R_a\) & 0 & 0 & 0 \\
Fully archived sessions & 3 & 3 & 3 \\
Archived operator code edits & 0 & 0 & 0 \\
Archived quantum-logic corrections & 0 & 0 & 0 \\
\hline
\end{tabular}
\end{table}

Within the nine fully archived sessions, the trial record also marked every first-response answer artifact as grader-passing. Only one first notebook execution completed without a follow-up. In the other eight, execution stopped at an optional \texttt{qc.draw("mpl")} call because \texttt{pylatexenc} was missing. One permitted installation or rendering adjustment resolved each case. One HW2 grading environment additionally lacked \texttt{qiskit-aer}. These were environment and visualization issues, not errors in the quantum algorithm, expected result, output schema, or reflection. No archived session required operator editing of the generated code.

\subsection{Exploratory Hardware Runs Remained LLM-Assisted}

Table~\ref{tab:hardware} summarizes the earlier exploratory hardware artifacts. ChatGPT produced or assisted with both simulator and IBM Runtime workflows. In each real-device run, the theoretically expected bitstring remained dominant despite noise. Transpilation substantially increased depth, but the result remained interpretable.

\begin{table}[t]
\caption{Earlier exploratory IBM hardware results}
\label{tab:hardware}
\centering
\footnotesize
\begin{tabular}{lccc}
\hline
\textbf{Task} & \textbf{Backend} & \textbf{Expected count} & \textbf{Depth} \\
\hline
QPE & \texttt{ibm\_marrakesh} & 725/1024 (70.80\%) & 8$\rightarrow$145 \\
QFT & \texttt{ibm\_kingston} & 3709/4096 (90.55\%) & 10$\rightarrow$32 \\
DJ & \texttt{ibm\_kingston} & 3438/4096 (83.94\%) & 8$\rightarrow$39 \\
\hline
\end{tabular}
\end{table}

Hardware added backend names, job identifiers, transpiled circuits, and noisy counts that ChatGPT could not predict before execution. It did not eliminate the minimally engaged workflow: ChatGPT generated the code, the operator ran it, and the resulting data could be returned for interpretation. The exploratory evidence therefore indicates that hardware supplied execution evidence without eliminating substantial AI assistance.

\section{Why the Modifications Did Not Stop ChatGPT}

\subsection{Personalization Changed Parameters, Not the Task Class}

A deterministic seed prevents every student from receiving exactly the same numeric configuration. It does not necessarily create a new reasoning problem. ChatGPT could read the visible generator, specialize the circuit to the generated values, and produce a parameterized solution. Non-palindromic bitstrings and reversed maps exposed careless implementations, but they remained explicit, rule-based transformations within the model's capability.

HW1 illustrates the distinction. Its bit flips, measurement map, and expected count key were deterministic functions of student-visible data. Once ChatGPT parsed those functions, it could construct both the exact output and the required circuit. Personalization improved assignment variety and grading logistics but did not force independent student reasoning.

\subsection{Scaffolding and Predictability Weakened Execution Dependence}

In HW2, QFT followed by inverse QFT is the identity:
\begin{equation}
\mathrm{QFT}^{-1}\mathrm{QFT}\lvert x\rangle=\lvert x\rangle .
\end{equation}
Moreover, the student-facing notebook exposed the complete QFT and inverse-QFT routines, the expected recovery relationship, the measurement-map calculation, and generic reflection text. ChatGPT could therefore complete the package without independently deriving the algorithm and could know the ideal count key before simulator execution. Circuit depth and operation-count fields made the submission richer, but they did not prevent completion of the supplied workflow.

HW3 similarly disclosed the mathematical relation between the oracle mask and the Deutsch--Jozsa classification. A zero mask is constant; a nonzero linear mask is balanced. Constructing the corresponding controlled-NOT oracle is a standard translation from the supplied mask. The custom seed removed a single fixed answer, but it did not move the task outside the model's capability.

\subsection{Hidden Grading Is Not Hidden Problem Information}

Hidden deterministic references made the assignments autogradable and prevented a student from simply reading an instructor answer file. They did not prevent ChatGPT from deriving a correct answer from the public specification. The graders checked seeded configuration fields, expected bitstrings, ideal counts, dominant outcomes, and reflection presence or length. They did not semantically validate every submitted notebook circuit, and optional QPY export was not a mandatory gate. Consequently, grader acceptance established consistency with expected outputs, not independent authorship or understanding.

Machine-readable JSON also did not create resistance. LLMs are well suited to structured output when a schema or example is supplied. Likewise, generic or result-specific short reflections were generated fluently after ChatGPT had calculated or received the relevant values.

\subsection{Minor Failures Were Easily Repairable}

The observed weaknesses---missing packages, \texttt{pylatexenc}, outdated Runtime syntax, and occasional bit-order risk---were not robust barriers. They were ordinary debugging issues that a student could paste back into the same conversation. This is consistent with quantum-code studies showing large gains from execution feedback \cite{slim2026quanbenchplus,novo2026domain}. A design is not meaningfully resilient if one exact error message converts a near-solution into a complete submission without requiring conceptual intervention.

\section{Implications for Engineering Education}

The negative result does not imply that personalization, autograding, simulator work, or hardware laboratories should be abandoned. Each remains educationally and operationally valuable. The finding is that their presence in a take-home package should not be interpreted as evidence that the student independently performed the substantive work.

A human-centric response should distinguish \emph{AI-supported production} from \emph{verified understanding}. Students may be allowed to use AI for syntax, setup, or debugging while instructors directly assess whether they can explain and adapt the result. Practical options include:
\begin{itemize}
    \item a short oral defense tied to the student's own circuit and counts;
    \item a supervised modification of a qubit mapping, oracle, or input state;
    \item prediction of a new outcome before execution;
    \item a transfer task that changes the algorithmic structure rather than only the seed;
    \item semantic validation of submitted circuits when implementation is a learning outcome;
    \item brief in-class checks that connect code, mathematics, and observed hardware noise.
\end{itemize}
These measures are recommendations derived from the observed failure mode; they were not experimentally compared in this study.

The result also cautions against equating more submission artifacts with stronger validity. A notebook, JSON file, transpiler statistics, QPY circuit, job identifier, and reflection can all be genuine artifacts while still being assembled through an LLM-directed workflow. Assessment should therefore specify which claim each artifact supports: execution occurred, a circuit is semantically correct, or the student understands and can transfer the concept. Those are different evidentiary claims.

\section{Limitations}

This is an exploratory negative-result study, not proof that every possible Qiskit assignment is solvable. Three fixed student-visible instances, one per implemented package, were tested at the introductory undergraduate level. Each instance was repeated in 50 separate ChatGPT sessions. This design supports a strong claim about consistency across separate sessions for those three instances, but not about every possible seed, assignment variant, more open-ended design task, private dynamic challenge, mandatory semantic circuit validator, adversarial noisy simulation, or supervised component.

All 150 execution and grader outcomes were verified, but complete transcripts, generated versions, logs, and grader output were retained for only nine sessions. The aggregate success result is therefore verified for all 150 sessions, while exact process-level replay is limited to the nine fully archived sessions.

The study focuses on ChatGPT. Exact model labels and dates were preserved for the nine archived sessions but not uniformly across the other sessions. Rapid model and Qiskit changes limit temporal reproducibility. No human participants were studied, so the work does not measure actual student behavior, learning, motivation, or over-reliance. There was also no randomized baseline or ablation study isolating the effect of each deterrence layer. Finally, optional hardware results demonstrate the feasibility of an LLM-assisted real-device workflow but were not part of the 150-session fixed-instance experiment.

\section{Conclusion}

The project began by trying to construct AI-resilient and autogradable Qiskit homework without banning AI. The central outcome was negative. For each of three fixed assignment instances, exactly 50 separate ChatGPT sessions were conducted. All 150 final artifacts executed and passed the corresponding grader after ChatGPT received the student-visible text or files. Nine fully archived sessions provide complete process-level provenance for the same successful outcome.

Seeded personalization, non-palindromic data, reversed measurement maps, custom oracle masks, hidden references, machine-readable output, circuit metrics, reflections, and optional hardware made the assignments richer and more verifiable. In the tested designs, they did not prevent direct ChatGPT completion. The appropriate conclusion is not that every possible quantum-programming assessment is futile. It is that correctness of a take-home artifact, even a personalized and execution-oriented one, is insufficient evidence of independent understanding. Engineering educators should design assessment around what students must explain, modify, predict, and defend, rather than assuming that a completed notebook establishes who performed the intellectual work.


\end{document}